\documentclass[conference]{IEEEtran}
\IEEEoverridecommandlockouts
\usepackage{cite}
\usepackage{amsmath,amssymb,amsfonts}
\usepackage{bm}
\usepackage[pdftex]{graphicx}

\usepackage{textcomp}
\usepackage{xcolor}

\newcommand{\Hline}{\noalign{\hrule height 0.4mm}}

\newcommand{\alphaijm}{\tilde{\alpha}_{ijm}}

\newcommand{\Bim}{\bm{B}_{im}}
\newcommand{\chiijm}{\chi_{ijm}}

\newcommand{\Em}{\bm{e}_m}

\newcommand{\etaijkmn}{\tilde{\eta}_{ijkmn}}

\newcommand{\ginm}{g_{inm}}

\newcommand{\Gin}{\bm{G}_{in}}
\newcommand{\Gindiag}{\bm{\mathcal{G}}_{in}}

\newcommand{\Him}{\bm{H}_{im}}
\newcommand{\Hp}{^\mathsf{H}}

\newcommand{\lijm}{l_{ijm}}

\newcommand{\phiijm}{\phi_{ijm}}
\newcommand{\Qi}{\bm{Q}_{i}}

\newcommand{\qim}{\bm{q}_{im}}

\newcommand{\rijm}{r_{ijm}}

\newcommand{\sigmaijn}{\sigma_{ijn}}

\newcommand{\sijnhat}{\hat{\bm{s}}_{ijn}}

\newcommand{\sumi}{\sum_i}

\newcommand{\sumj}{\sum_j}
\newcommand{\sumn}{\sum_n}
\newcommand{\summ}{\sum_m}
\newcommand{\sumk}{\sum_k}
\newcommand{\sumij}{\sum_{i,j}}
\newcommand{\sumkn}{\sum_{k,n}}
\newcommand{\sumijm}{\sum_{i,j,m}}
\newcommand{\tik}{t_{ik}}

\newcommand{\Uim}{\bm{U}_{im}}
\newcommand{\Tp}{^\mathsf{T}}

\newcommand{\vkj}{v_{kj}}

\newcommand{\xiijm}{\tilde{\xi}_{ijm}}

\newcommand{\x}{\bm{x}}
\newcommand{\xij}{\bm{x}_{ij}}

\newcommand{\Xijhat}{\hat{\bm{X}}_{ij}}

\newcommand{\zetaijm}{\tilde{\zeta}_{ijm}}

\newcommand{\zkn}{z_{kn}}

\newcommand{\matnn}[9]{
\begin{bmatrix}
    #1     & #2     & \cdots & #3 \\
    #4     & #5     & \cdots & #6 \\
    \vdots & \vdots & \ddots & \vdots \\
    #7     & #8     & \cdots & #9
\end{bmatrix}
}

\def\BibTeX{{\rm B\kern-.05em{\sc i\kern-.025em b}\kern-.08em
    T\kern-.1667em\lower.7ex\hbox{E}\kern-.125emX}}
\begin{document}
\title{
Joint-Diagonalizability-Constrained Multichannel Nonnegative Matrix Factorization Based on Multivariate Complex Sub-Gaussian Distribution
\thanks{
  This work was partly supported by SECOM Science and TechnologyFoundation and JSPS KAKENHI Grant Numbers JP19H01116, JP19H04131, and JP19K20306, and JSPS-CAS Joint Research Program, Grant number JPJSBP120197203.
}}

\author{
    \IEEEauthorblockN{
        Keigo Kamo\IEEEauthorrefmark{1},
        Yuki Kubo\IEEEauthorrefmark{1},
        Norihiro Takamune\IEEEauthorrefmark{1},
        Daichi Kitamura\IEEEauthorrefmark{2}\\
        Hiroshi Saruwatari\IEEEauthorrefmark{1},
        Yu Takahashi\IEEEauthorrefmark{3},
        Kazunobu Kondo\IEEEauthorrefmark{3}
    }
    \IEEEauthorblockA{
        \IEEEauthorrefmark{1}
        Graduate School of Information Science and Technology, The University of Tokyo,
        Tokyo 113-8656, Japan
    }
    \IEEEauthorblockA{
        \IEEEauthorrefmark{2}
        National Institute of Technology, Kagawa College,
        Kagawa 761-8058, Japan
    }
    \IEEEauthorblockA{
        \IEEEauthorrefmark{3}
        R\&D Division, Yamaha Corporation,
        Shizuoka 430-8650, Japan
    }
}

\maketitle

\begin{abstract}
In this paper, we address a statistical model extension of multichannel nonnegative matrix factorization (MNMF)
for blind source separation, and we propose a new parameter update algorithm used in the sub-Gaussian model.
MNMF employs full-rank spatial covariance matrices and can simulate situations in which the reverberation is strong
and the sources are not point sources.
In conventional MNMF, spectrograms of observed signals are assumed to follow a multivariate Gaussian distribution.
In this paper, first, to extend the MNMF model, we introduce the multivariate generalized Gaussian distribution
as the multivariate sub-Gaussian distribution. Since the cost function of MNMF based on this multivariate sub-Gaussian model is difficult to minimize,
 we additionally introduce the joint-diagonalizability constraint in spatial covariance matrices to MNMF similarly to FastMNMF,
and transform the cost function to the form to which we can apply the auxiliary functions to
derive the valid parameter update rules.
Finally, from blind source separation experiments,
we show that the proposed method outperforms the conventional methods in source-separation accuracy.
\end{abstract}


\vspace{-4pt}
\section{Introduction\label{sec:intro}}
\vspace{-1pt}
Blind source separation (BSS)~\cite{sawada2019review} is a technique to separate sound sources from observed mixtures without any prior information about the sources or mixing system. 
For a determined or overdetermined situation,
when the sources are point sources and the reverberation is sufficiently short (referred to as  the \textit{rank-1 spatial model}),
frequency-domain independent component analysis~\cite{smaragdis1998blind,saruwatari2006blind}, independent vector analysis (IVA)~\cite{hiroe2006solution,kim2006independent,kim2006blind}, and independent low-rank matrix analysis (ILRMA)~\cite{kitamura2016determined,kitamura2018determined}
have been proposed.
However, the rank-1 spatial model does not hold in the case of spatially spread sources or strong reverberation.

Multichannel nonnegative matrix factorization (MNMF)~\cite{ozerov2010multichannel,sawada2013multichannel}
is an extension of nonnegative matrix factorization (NMF)~\cite{lee1999learning} to multichannel cases, which estimates the spatial covariance matrices of each source.
MNMF employs full-rank spatial covariance matrices~\cite{duong2010under}, and this model can
simulate situations where, e.g., the reverberation is longer than the length of time-frequency analysis.
However, it has been reported that MNMF has a huge computational cost
and its performance strongly depends on the initial values of parameters~\cite{kitamura2016determined}.

In the original MNMF,
the observed signal is assumed to follow a time-variant multivariate complex Gaussian distribution.
Recently, the generative model extension of MNMF to multivariate complex Student's \textit{t} distribution (\textit{t}-MNMF~\cite{kitamura2016student}) has been proposed.
However, the original Gaussian MNMF and \textit{t}-MNMF cannot assume that the generative model follows a multivariate sub-Gaussian distribution,
whereas we reported that the separation accuracy is improved by expanding the source signal model
to the univariate sub-Gaussian distribution in ILRMA (hereafter referred to as sub-Gaussian ILRMA)~\cite{mogami2019independent}.
Thus, we can expect that the introduction of sub-Gaussianity in MNMF leads to the improvement of separation accuracy.

In this paper, we provide three contributions, namely,
a generalization of the generative model,
its parameter optimization algorithm with a joint-diagonalizability constraint,
and experimental evaluation of the proposed MNMF.
First,
we extend the generative model of MNMF to a time-variant multivariate complex sub-Gaussian distribution;
this is hereafter referred to as \textit{sub-Gaussian MNMF}.
We employ a multivariate complex generalized Gaussian distribution (GGD) as a sub-Gaussian distribution
by restricting its shape parameter.
It is reported that some musical instrument signals obey sub-Gaussian distributions~\cite{naik2012audio}.
Next,
we derive parameter update rules of sub-Gaussian MNMF.
The cost function of sub-Gaussian MNMF is difficult to minimize, and the auxiliary function for the majorization-minimiazation algorithm~\cite{hunter2000quantile} has not been discovered so far.
To design the valid auxiliary function, we introduce a joint-diagonalizability constraint in spatial covariance matrices.
This constraint has been introduced for the first time in FastMNMF~\cite{ito2019fastmnmf} to reduce the computational complexity of MNMF.
On the other hand, in this paper, we employ this joint-diagonalizability constraint
not to reduce the computational complexity
but to transform the cost function to the form to which we can apply the auxiliary function technique.
To the best of our knowledge, this is the world's first update algorithm of parameters
that guarantees a monotonic nonincrease in the cost function
of MNMF with a multivariate complex sub-Gaussian model.
Finally,
we conduct BSS experiments under reverberant conditions,
showing that the proposed sub-Gaussian MNMF outperforms conventional methods in source-separation accuracy.

\section{Conventional Methods}
\subsection{MNMF~\cite{ozerov2010multichannel,sawada2013multichannel}\label{sub:MNMF}}
The short-time Fourier transform (STFT) of the observed multichannel signal is defined as
\begin{align}
    \label{form:gauss_dist}
    \xij = (x_{ij,1},\dots,x_{ij,M})\Tp \in \mathbb{C}^M,
  \end{align}
where $i=1,\dots,I,\,j=1,\dots,J,$ and $m=1,\dots,M$ are the indices
of the frequency bins, time frames, and channels, respectively, and $\cdot\Tp$ denotes the transpose.
The MNMF model estimates
a time-variant parameter $\sigmaijn$, which represents
a character of the source, and
a time-invariant parameter $\Gin$, which represents
spatial characteristics of the source, where $n=1,\dots,N$ is the index of the sources.
$\sigmaijn$ corresponds to a power spectrogram and
$\Gin$ is called a spatial covariance matrix.
In MNMF, as the generative model of the observed signal $\xij$,
the following multivariate complex Gaussian distribution is assumed:
\begin{align}
  \label{form:generativemodel}
  \xij \sim \mathcal{N}\Bigl(\mathbf{0}_{M},\sumn\sigmaijn\Gin\Bigr),
\end{align}
where $\mathbf{0}_M\in\mathbb{C}^M$ is an $M$-dimensional zero vector and
$\mathcal{N}(\bm{\mu},\bm{\Sigma})$ is the multivariate complex Gaussian distribution
whose mean is $\bm{\mu}$ and the covariance matrix is $\bm{\Sigma}$.
The source model $\sigmaijn$ is a spectrogram of the $n$th source
at the $i$th frequency and $j$th time frame,
having a low-rank spectral structure represented by NMF, as
\begin{align}
\label{form:powerspec}
  \sigmaijn=\sumk\tik\vkj\zkn,
\end{align}
where $k=1,\dots,K$ is the index of the NMF basis, and
$\tik\in\mathbb{R}_{\geq 0}$ and $\vkj\in\mathbb{R}_{\geq 0}$ represent
the $i$th frequency component of the $k$th basis
and the $j$th time-frame activation component of the $k$th basis, respectively.
In addition, $\zkn\in\mathbb{R}_{\geq 0}$ is a latent variable that indicates
whether the $k$th basis belongs to the $n$th source.
From \eqref{form:generativemodel},
the negative log-likelihood of the observed signal,
which is a cost function to be minimized, is given by
\begin{align}
  \label{cost:MNMF}
  \mathcal{L}_{\mathrm{MNMF}} &\overset{c}{=} \sumij\Bigl(\xij\Hp\Xijhat^{-1}\xij+\log\det\Xijhat\Bigr),
\end{align}
where $\Xijhat=\sumn\sigmaijn\Gin=\sumkn\tik\vkj\zkn\Gin$ and
$\overset{c}{=}$ denotes equality up to a constant.
We can estimate $\tik,\vkj,\zkn,$ and $\Gin$ by minimizing \eqref{cost:MNMF}.
$\Gin$ can be optimized by solving the Riccati equation and
the remaining parameters are updated by using the auxiliary function technique
(details of these update rules are described in \cite{sawada2013multichannel}).
After the update, we can estimate the separated signal $\sijnhat$ using the multichannel Wiener filter:
\begin{align}
\label{formula:multichannelwienr}
    \sijnhat = \biggl(\sumk\tik\vkj\zkn\Gin\biggr)\Xijhat^{-1}\xij.
\end{align}
MNMF assumes that $\Gin$ is a full-rank matrix~\cite{duong2010under}, and this increases versatility for
various types of spatial condition. However, this full-rank nature requires a large
amount of computation.

\subsection{FastMNMF~\cite{ito2019fastmnmf,sekiguchi2019fast}}
To reduce the computational complexity of the update algorithm, FastMNMF additionally assumes that the spatial covariance matrices
$\boldsymbol{G}_{i1},\dots,\boldsymbol{G}_{iN}$ are jointly diagonalizable
by $\Qi=(\boldsymbol{q}_{i1},\dots,$ $\boldsymbol{q}_{iM})\Hp$, which does not depend on the source index $n$, as
\begin{align}
  \label{formula:jointdiag}
  \Qi\Gin\Qi\Hp = \Gindiag \hspace{5mm} (n=1,\dots,N),
\end{align}
where $\Gindiag$ is a diagonal matrix.
From (\ref{cost:MNMF}) and (\ref{formula:jointdiag}),
the negative log-likelihood of the observed signal is given by
\begin{align}
\label{cost:FastMNMF}
    \mathcal{L}_{\mathrm{F}} &\overset{c}{=} \sumijm\biggl[\frac{|\qim\Hp\xij|^2}{\sum_{n,k}\tik\vkj\zkn\ginm} + \log\sum_{n,k}\tik\vkj\zkn\ginm\biggr] \nonumber\\
                &\quad -2J\sumi\log|\det\Qi|,
\end{align}
where $\ginm$ is the $m$th diagonal element of $\Gindiag$.
The joint-diagonalization matrix $\Qi$ in (\ref{cost:FastMNMF}) can be optimized by iterative projection (IP)~\cite{ono2011stable}, and the remaining parameters
are updated by using the auxiliary function technique~\cite{sekiguchi2019fast}.
Note that the algorithm of the update of spatial covariance matrices in FastMNMF is different from that in MNMF, which uses the Riccati equation.
This algorithmic improvement provides a different separation performance; FastMNMF is almost the same as or slightly better than MNMF~\cite{sekiguchi2019fast, kubo2019efficient}.
After the update, we can estimate the separated signal using the multichannel Wiener filter similarly to \eqref{formula:multichannelwienr}.

\section{Proposed Method\label{sec:proposed}}
\subsection{Motivation and Strategy\label{subsec:motivation}}
In this paper,
we propose a model extension of MNMF to the multivariate complex sub-Gaussian distribution.
Sub-Gaussian MNMF can appropriately model the observed signal that follows the sub-Gaussian distribution, which cannot be
represented  by the conventional methods.
The multivariate GGD is used as the sub-Gaussian distribution.
The probability density function of the zero-mean multivariate GGD is given by
\begin{align}
  \label{form:GGD}
  p(\x; \bm{0}_{M},\bm{\Sigma},\beta) = \frac{C(\beta)}{\det\bm{\Sigma}}\exp\Bigl(-(\x\Hp\bm{\Sigma}^{-1}\x)^{\beta/2}\Bigr),
\end{align}
where $\beta>0$ is the shape parameter and
$C(\beta)$ is the normalizing constant of the multivariate GGD.
In the case of $0<\beta<2$, the GGD becomes super-Gaussian.
In the case of $\beta>2$, the GGD becomes sub-Gaussian.
For $\beta= 2$, the GGD corresponds to the Gaussian distribution.
The negative log-likelihood of MNMF based on the multivariate GGD model is represented as
\begin{align}
  \label{cost:GGD-MNMF}
  \mathcal{L}_{\mathrm{GGD}} \overset{c}{=} \sumij\Bigl((\xij\Hp\Xijhat^{-1}\xij)^{\beta/2}+\log\det\Xijhat\Bigr),
\end{align}
where we substitute the model parameter $\Xijhat$ for $\Sigma$ in \eqref{form:GGD}.

In the case of $\beta>2$, this cost function \eqref{cost:GGD-MNMF} is difficult to minimize, especially in the parameter $\Gin$
because it is impossible to design a majorization function to which we can apply the Riccati equation solver
(no quadratic function can majorize the first term of the right-hand side of \eqref{cost:GGD-MNMF}).
Thus, we cannot derive the update rules of the sub-Gaussian MNMF that guarantees a monotonic nonincrease in the cost function
without any constraint.
Hence, we additionally introduce the joint-diagonalizability constraint.
If we apply the joint-diagonalizability constraint and an appropriate auxiliary function to \eqref{cost:GGD-MNMF},
we can find that the optimization problem here is identical to a demixing-matrix-optimization problem in sub-Gaussian ILRMA,
which can be solved by the authors' previous work, \textit{generalized IP}~\cite{mogami2019independent}.
In the next subsections, details of the proposed parameter update algorithm is described.

\subsection{Proposed Sub-Gaussian MNMF\label{subsec:proposed}}
By substituting \eqref{formula:jointdiag} into \eqref{cost:GGD-MNMF},
we obtain the cost function of MNMF with the multivariate GGD model as
\begin{align}
  \nonumber
  \label{cost:GGD-FastMNMF}
  \mathcal{L}_{\mathrm{GGD}} &\overset{c}{=} -2J\sumi\log|\det\Qi| + \sumijm\log\sum_{n,k}\tik\vkj\zkn\ginm \\
                             &\phantom{=} +\sumij\biggl(\summ\frac{|\qim\Hp\xij|^2}{\sum_{n,k}\tik\vkj\zkn\ginm}\biggr)^{\frac{\beta}{2}}.
\end{align}
First, we derive update rules of parameters $\tik,\vkj,\zkn,$ and $\ginm$.
For $\beta>2$, $f_1(y)=y^{\beta/2}$ is a convex function.
Hence, by applying Jensen's inequality to \eqref{cost:GGD-FastMNMF}, we have
\begin{align}
  \nonumber
  \mathcal{L}_{\mathrm{GGD}}  &\overset{c}{=} \sumij\biggl(\summ\frac{|\qim\Hp\xij|^2}{\sum_{n,k}\tik\vkj\zkn\ginm}\biggr)^{\frac{\beta}{2}} \\
  \nonumber
                              &\phantom{=}+\sumijm\log\sum_{n,k}\tik\vkj\zkn\ginm \\
                              \nonumber
                              \label{cost:auxiliary1}
                              &\overset{c}{\leq} \sumij\summ\xiijm\biggl(\frac{|\qim\Hp\xij|^2}{\xiijm\sum_{n,k}\tik\vkj\zkn\ginm}\biggr)^{\frac{\beta}{2}} \\
                              &\phantom{=}+\sumijm\log\sum_{n,k}\tik\vkj\zkn\ginm,
\end{align}
where $\xiijm\geq0$ is an auxiliary variable that satisfies $\summ\xiijm=1$, and
$\overset{c}{\leq}$ denotes that the left-hand side is less than or equal to the right-hand side up to a constant.
Moreover, we can design an auxiliary function of $f_2(y)=(1/y)^{\beta/2}$ by applying Jensen's inequality because it is also a convex function.
We can also design one of $f_3(y)=\log y$ by applying the tangent inequality because it is a concave function.
Then, we can obtain the final auxiliary function $\mathcal{L}^+_{\mathrm{Sub}}$ as
\begin{align}
  \nonumber
  \label{cost:auxiliary_tvzg}
  \mathcal{L}_{\mathrm{GGD}} &\overset{c}{\leq} \sumijm(\xiijm)^{1-\frac{\beta}{2}}\sum_{n,k}\etaijkmn\biggl(\frac{\etaijkmn|\qim\Hp\xij|^2}{\tik\vkj\zkn\ginm}\biggr)^{\frac{\beta}{2}} \\
                             &\phantom{=}+\sumijm\frac{1}{\zetaijm}\sum_{n,k}\tik\vkj\zkn\ginm \\
                             &:=\mathcal{L}^+_{\mathrm{Sub}},
\end{align}
where $\etaijkmn\geq0, \zetaijm\geq0$ are auxiliary variables, and $\zetaijm$ satisfies $\summ\zetaijm=1$.
The equalities of \eqref{cost:auxiliary1} and \eqref{cost:auxiliary_tvzg} hold if and only if the auxiliary variables are set as follows:
\begin{align}
  \label{eq:xiijm}
  \xiijm &= \left. \frac{|\qim\Hp\xij|^2}{\sumkn\tik\vkj\zkn\ginm} \middle/ \sum_{m'}\frac{|\bm{q}_{im'}\Hp\xij|^2}{\sumkn\tik\vkj\zkn g_{inm'}} \right., \\
  \etaijkmn &= \frac{\tik\vkj\zkn\ginm}{\sumkn\tik\vkj\zkn\ginm},\\
  \label{eq:zetaijkmn}
  \zetaijm &= \sumkn\tik\vkj\zkn\ginm.
\end{align}
The update rules for \eqref{cost:auxiliary_tvzg} w.r.t. $\tik,\vkj,\zkn,$ and $\ginm$ are derived by setting the gradient to zero.
From 
$\partial\mathcal{L}_{\mathrm{Sub}}^+/\partial\tik=0$,
$\partial\mathcal{L}_{\mathrm{Sub}}^+/\partial\vkj=0$,
$\partial\mathcal{L}_{\mathrm{Sub}}^+/\partial\zkn=0$,
and $\partial\mathcal{L}_{\mathrm{Sub}}^+/\partial\ginm=0$,
we obtain
\begin{align}
  \label{chiijm}
  \chiijm &= \sumkn\tik\vkj\zkn\ginm, \\
  \phiijm &= |\qim\Hp\xij|^2\biggl(\sum_{m'}\frac{|\bm{q}_{im'}\Hp\xij|^2}{\chi_{ijm'}}\biggr)^\frac{\beta-2}{2}, \\
  \label{formula:update_GGD_t}
  \tik &\leftarrow \tik \Biggl(\frac{\beta\sum_{j,n,m}\frac{\phiijm\vkj\zkn\ginm}{\chiijm^2}}{2\sum_{j,n,m}\frac{\vkj\zkn\ginm}{\chiijm}}\Biggr)^{\frac{2}{\beta+2}}, \\
  \vkj &\leftarrow \vkj \Biggl(\frac{\beta\sum_{i,n,m}\frac{\phiijm\tik\zkn\ginm}{\chiijm^2}}{2\sum_{i,n,m}\frac{\tik\zkn\ginm}{\chiijm}}\Biggr)^{\frac{2}{\beta+2}}, \\
  \zkn &\leftarrow \zkn \Biggl(\frac{\beta\sum_{i,j,m}\frac{\phiijm\tik\vkj\ginm}{\chiijm^2}}{2\sum_{i,j,m}\frac{\tik\vkj\ginm}{\chiijm}}\Biggr)^{\frac{2}{\beta+2}}, \\
  \label{formula:update_GGD_g}
  \ginm &\leftarrow \ginm\Biggl(\frac{\beta\sum_{j,k}  \frac{\phiijm\tik\vkj\zkn }{\chiijm^2}}{2\sum_{j,k}  \frac{\tik\vkj\zkn }{\chiijm}}\Biggr)^{\frac{2}{\beta+2}}.
\end{align}
Note that these update rules have already been substituted under the equality conditions \eqref{eq:xiijm}--\eqref{eq:zetaijkmn} and rearranged.

Next,
we derive the update rule of the joint-diagonalization matrix $\Qi$.
We can design the auxiliary function of the cost function \eqref{cost:GGD-FastMNMF} w.r.t. $\Qi$
similarly to the derivation of \eqref{cost:auxiliary1} as
\begin{align}
  \nonumber
  \mathcal{L}_{\mathrm{GGD}} &\overset{c}{\leq} \sumijm\xiijm\biggl(\frac{|\qim\Hp\xij|^2}{\xiijm\sum_{n,k}\tik\vkj\zkn\ginm}\biggr)^{\frac{\beta}{2}} \\
  \nonumber
                             &\phantom{=} - 2J\sumi\log|\det\Qi| \\
                             \label{cost:auxiliary2}
                             &= \sumijm\frac{|\qim\Hp\xij|^\beta}{\rijm^\beta} - 2J\sumi\log|\det\Qi|,
\end{align}
where
\begin{align}
  \rijm = \xiijm^{\frac{1}{2}-\frac{1}{\beta}}\biggl(\sumkn\tik\vkj\zkn\ginm\biggr)^{\frac{1}{2}}.
\end{align}
The auxiliary function \eqref{cost:auxiliary2}
is the same form as that for the optimization of a demixing matrix in sub-Gaussian ILRMA~\cite{mogami2019independent}.
Hence, we derive the update rule in the same manner as sub-Gaussian ILRMA.
Similarly to \cite{mogami2019independent}, in the case of $2<\beta\leq4$,
we can bound the term $|\qim\Hp\xij|^\beta$ using the inequality of weighted arithmetic and geometric means as follows:
\begin{align}
  \label{formula:WAGmean}
  |\qim\Hp\xij|^\beta \leq \frac{\beta}{4}\frac{|\qim\Hp\xij|^4}{\alphaijm^{4-\beta}}+\Bigl(1-\frac{\beta}{4}\Bigr)\alphaijm^\beta,
\end{align}
where $\alphaijm$ is an auxiliary variable and the equality of \eqref{formula:WAGmean} holds if and only if $\alphaijm=|\qim\Hp\xij|$.
We can apply \eqref{formula:WAGmean} to \eqref{cost:auxiliary2} and obtain
\begin{align}
  \label{cost:auxiliary3}
  \mathcal{L}_{\mathrm{GGD}} \overset{c}{\leq} \frac{\beta}{4}\sumijm\frac{|\qim\Hp\xij|^4}{\alphaijm^{4-\beta}\rijm^\beta} - 2J\sumi\log|\det\Qi|.
\end{align}
Moreover,
we can design a further auxiliary function of \eqref{cost:auxiliary3} as
\begin{align}
  \label{cost:auxiliary4}
  \mathcal{L}_{\mathrm{GGD}} \leq J\sum_{i,n}(\qim\Hp\Bim\qim)^2 - 2J\sumi\log|\det\Qi|,
\end{align}
where $\Bim$ is an auxiliary variable that satisfies
\begin{align}
  \lijm       &= \sqrt[4]{|\qim\Hp\xij|^{4-\beta}\rijm^\beta}, \\
  \Him        &= \biggl[\frac{1}{l_{i1m}}\bm{x}_{i1}\cdots\frac{1}{l_{iJm}}\bm{x}_{iJ}\biggr], \\
  \bm{a}_{im} &= \biggl[ a_{i1m} \cdots a_{iJm} \biggr]\Tp = \Him\Hp\tilde{\bm{q}}_{im}, \\
  \bm{A}_{im} &= \matnn{||\bm{a}_{im}||^2}{-a_{i1m}\overline{a_{i2m}}}{-a_{i1m}\overline{a_{iJm}}}{-a_{i2m}\overline{a_{i1m}}}{||\bm{a}_{im}||^2}{-a_{i2m}\overline{a_{iJm}}}{-a_{iJm}\overline{a_{i1m}}}{-a_{iJm}\overline{a_{i2m}}}{||\bm{a}_{im}||^2},\\
  \Bim  &= \frac{\sqrt{\beta}}{2\sqrt{J\sumj|a_{ijm}|^4}}\Him\bm{A}_{im}\Him\Hp.
 \end{align}
Here, $\overline{\cdot}$ denotes the complex conjugate
and $\tilde{\bm{q}}_{im}$ is an auxiliary variable and the equality of \eqref{cost:auxiliary4} holds if and only if $\tilde{\bm{q}}_{im}=\qim$.
The joint-diagonalization matrix $\Qi$ of the cost function \eqref{cost:auxiliary4} can be updated by generalized IP~\cite{mogami2019independent}.
The update rule of $\Qi$ can be derived as
\begin{align}
  \label{formula:update_GGD_l}
  \rijm &\leftarrow |\qim\Hp\xij|^{1-\frac{2}{\beta}}\chiijm^{\frac{1}{\beta}} (\sum_{m'}\frac{|\bm{q}_{im'}\Hp\xij|^2}{\chi_{ijm'}})^{\frac{1}{\beta}-\frac{1}{2}}\\ 
  \Uim  &= \sumj\frac{\xij\xij\Hp}{\sqrt{|\qim\Hp\xij|^{4-\beta}\rijm^\beta}},
\end{align}
\begin{align}
  \nonumber
  \Bim'  &= \qim\Hp\Uim\qim\Uim +\sumj\frac{|\qim\Hp\xij|^{\beta-2}}{\rijm^\beta}\xij\xij\Hp\\
         &\phantom{=} -\Bigl(\Uim\qim\Bigr)\Bigl(\Uim\qim\Bigr)\Hp, \\
  \qim &\leftarrow (\Qi\Bim')^{-1}\Em, \\
  \label{formula:update_GGD_q_scale}
  \qim  &\leftarrow \qim \biggl(\frac{2J}{\beta\sumj(|\qim\Hp\xij|^\beta/\rijm^\beta)}\biggr)^{\frac{1}{\beta}},
\end{align}
where $\chiijm$ is defined in \eqref{chiijm}, and $\Bim'$ is equal to $\Bim$ up to scale.
From \eqref{formula:update_GGD_t}--\eqref{formula:update_GGD_g} and \eqref{formula:update_GGD_l}--\eqref{formula:update_GGD_q_scale},
all of the parameters $\tik,\vkj,\zkn,\ginm$, and $\Qi$ can be iteratively updated in the proposed sub-Gaussian MNMF.


\section{Experiments\label{sec:experiments}}
\subsection{Experimental Conditions\label{subsec:conditions}}
\begin{figure}[tb]
    \centering
    \includegraphics[width=8.5cm]{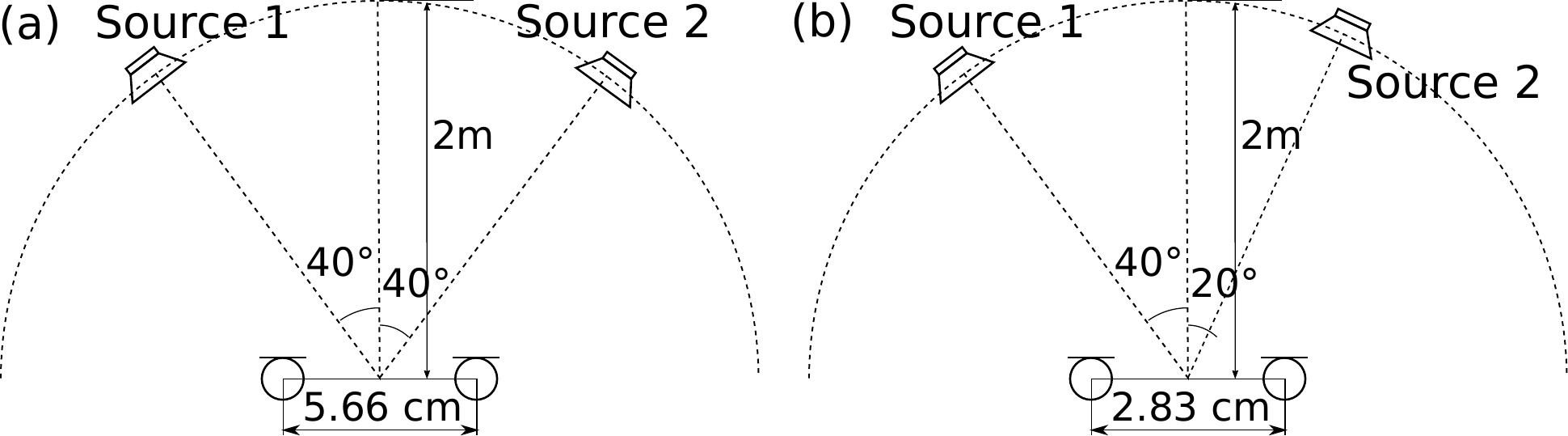}
    \caption{Spatial arrangements of sources and microphones.\label{mic}}
\end{figure}

We confirmed the efficacy of the proposed method by conducting 
music source separation experiments.
We compared six methods:
IVA~\cite{ono2011stable},
ILRMA~\cite{kitamura2016determined},
sub-Gaussian ILRMA~\cite{mogami2019independent},
MNMF~\cite{sawada2013multichannel},
FastMNMF~\cite{ito2019fastmnmf}, and
the proposed sub-Gaussian MNMF.
We used monaural dry music sources
of four melody parts~\cite{kitamura2015multichannel, kitamura2018open}.
Eight combinations of instruments with different melody parts were selected as shown in Table~\ref{table:music_dry_sources}.
\begin{table}[t]
  \caption{Dry sources used in experiment\label{table:music_dry_sources}}
  \centering
  \begin{tabular}{c|c|c}
  \Hline
          & Part name & Source (1/2) \\ \hline
  Music 1 & Midrange/Melody 2 & Flute/Piano \\
  Music 2 & Melody 1/Melody 2 & Flute/Oboe \\
  Music 3 & Melody 2/Midrange & Harpsichord/Violin \\
  Music 4 & Melody 2/Bass     & Cello/Violin \\
  Music 5 & Melody 1/Bass     & Cello/Oboe \\
  Music 6 & Melody 2/Melody 1 & Trumpet/Violin \\
  Music 7 & Bass/Melody 2     & Flute/Bassoon \\
  Music 8 & Bass/Melody 1     & Trumpet/Bassoon \\
  \Hline
  \end{tabular}
\end{table}
To simulate reverberant mixing, two-channel mixed signals were produced by convoluting the impulse response E2A ($T_{60}=$ 300~ms) in the RWCP database~\cite{nakamura2000acoustical}.
Fig.~\ref{mic} shows the recording conditions of E2A used in our experiments.
In these mixtures, the input signal-to-noise ratio was 0~dB.
The sampling frequency was 16~kHz and an STFT was performed using a 64~ms Hamming window with a 16~ms shift ($T_{60}$ was longer than the window length, i.e., the spatial covariance
matrices were full rank).
The total number of bases in the low-rank source model was $K=20, 30, 40, 50,$ and $ 60$.
The initializations of the source model parameters ($\tik,\vkj,\zkn$) and the spatial covariance matrix $\Gin$
were random values and the identity matrix, respectively.
The initialization of $\Qi$ was the identity matrix.
The shape parameter $\beta$ in sub-Gaussian ILRMA and the proposed sub-Gaussian MNMF was set to 4.
The number of iterations in all methods was 6000.
We used the source-to-distortion ratio (SDR) improvement~\cite{vincent2006performance} to evaluate the
total separation performance.

\subsection{Experimental Results\label{subsec:results}}
\begin{figure}[tb]
  \centering
  \includegraphics[width=8.5cm]{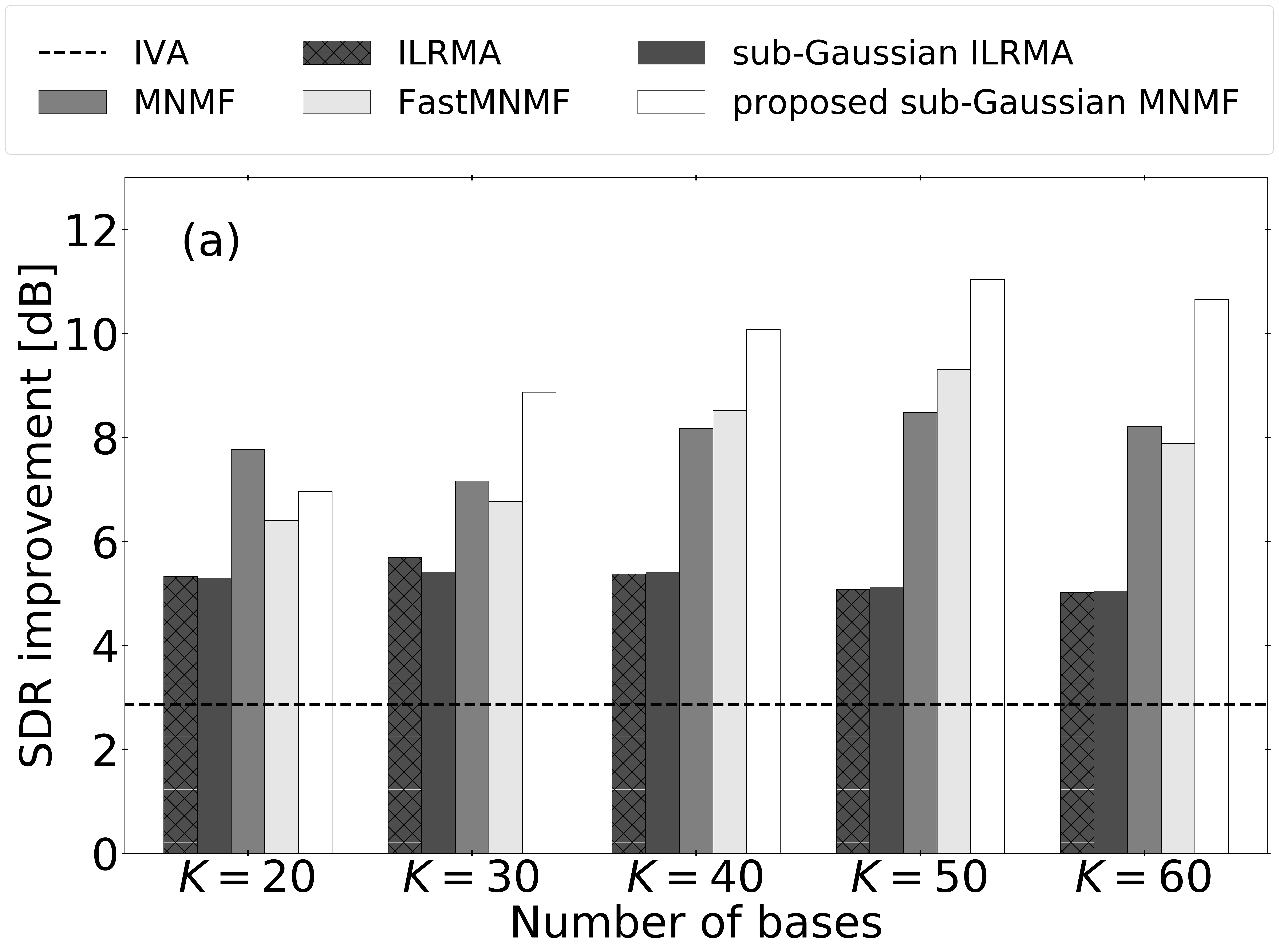}
  \vspace{-2mm}
\end{figure}
\begin{figure}[tb]
  \centering
  \includegraphics[width=8.5cm]{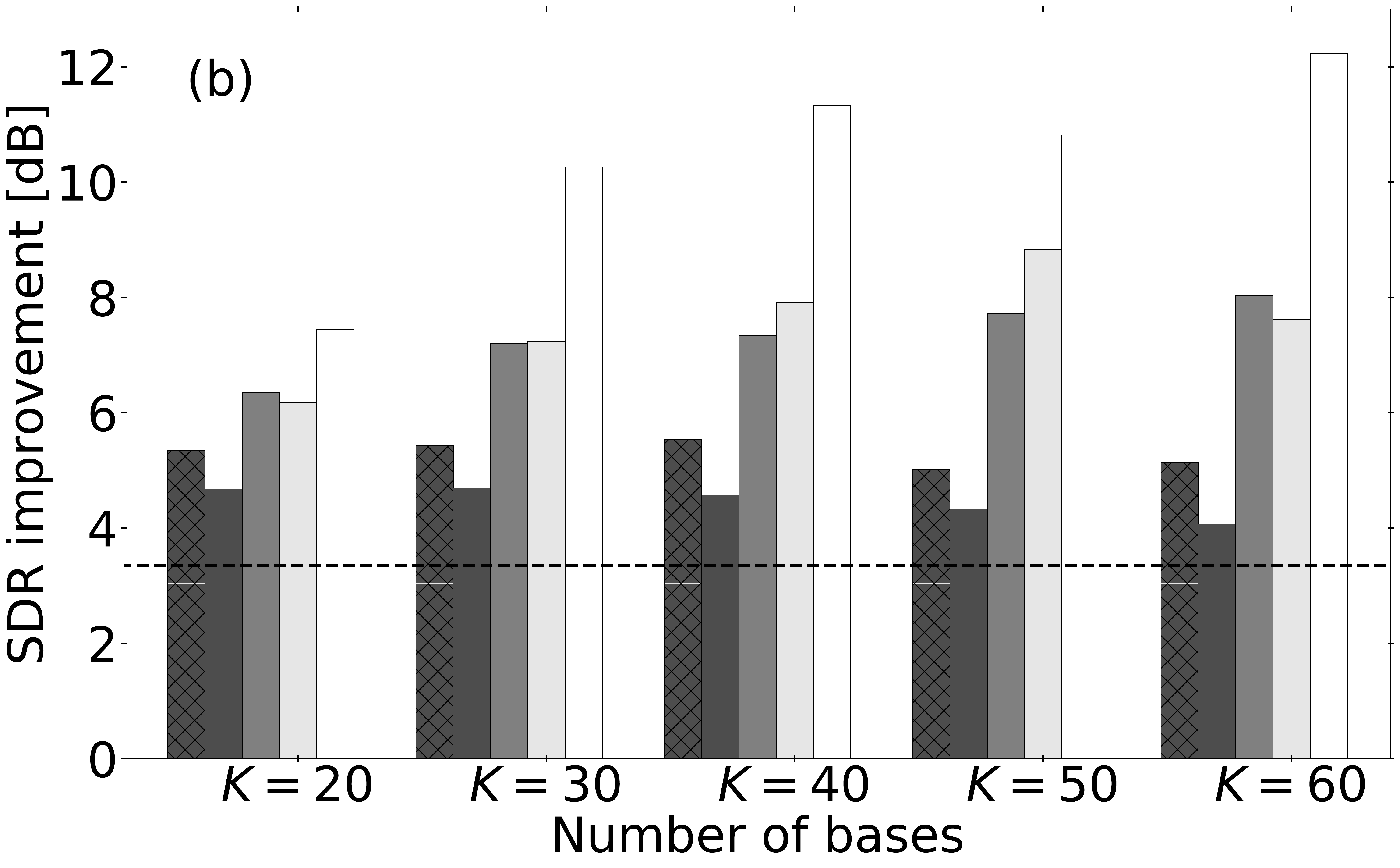}
  \vspace{-2mm}
  \caption{Resultant SDR improvement for each method. (a) Results of recording condition (a) in Fig. \ref{mic}. (b) Results
  of recording condition (b) in Fig. \ref{mic}.\label{results}}
\end{figure}
Fig.~\ref{results} shows the average SDR improvements over the source pairs and 10-trial initialization.
Compared with IVA, ILRMA, and sub-Gaussian ILRMA, both conventional MNMF and FastMNMF as well as the proposed sub-Gaussian
MNMF provide better SDR improvements.
This is because the rank-1 spatial model (IVA, ILRMA, and sub-Gaussian ILRMA) does not hold in the case of strong reverberation.
The conventional full-rank spatial models (MNMF and FastMNMF) achieve almost the same SDR improvements.
On the other hand, the proposed sub-Gaussian MNMF markedly outperforms the conventional full-rank spatial model methods, regardless of the spatial arrangement.
This suggests that the sub-Gaussian model is more appropriate than the conventional Gaussian model for simulating music signals,
showing the effectiveness of the proposed sub-Gaussian MNMF.

\section{conclusion}
In this paper, we proposed the model extension of MNMF to the multivariate complex sub-Gaussian distribution.
We introduced the joint-diagonalizability constraint in spatial covariance matrices
to derive update rules that guarantees a monotonic nonincrease in the cost function.
From the source-separation experiments,
we showed that the proposed method outperformed the conventional methods in terms of SDR improvement.

\bibliographystyle{IEEEbib}
\bibliography{refs}

\end{document}